**PySTACHIO: Python Single-molecule TrAcking stoiCHiometry Intensity and simulatiOn, a flexible, extensible, beginner-friendly and optimized program for analysis of single-molecule microscopy data**

Jack W Shepherd*[1,2], Ed J Higgins*[1,3], Adam J M Wollman[4], Mark C Leake‡[1,2]

[1] Department of Physics, University of York, York, YO10 5DD

[2] Department of Biology, University of York, York, YO10 5DD

[3] IT services, University of York, York, YO10 5DD

[4] Biosciences Institute, Newcastle University, Newcastle, NE1 7RU

‡ To whom correspondence should be addressed. E-mail mark.leake@york.ac.uk

*These authors contributed equally

**Highlights**

- We present PySTACHIO, a refined version of our spot tracking algorithm
- We demonstrate highly improved performance over previous MATLAB versions
- PySTACHIO can accurately estimate stoichiometries and 2D diffusion coefficients
- Performance is comparable to state-of-the-art packages on challenge data
- PySTACHIO has both GUI and command line interfaces and can be hosted as a web app

**Abstract**

As camera pixel arrays have grown larger and faster, and optical microscopy techniques ever more refined, there has been an explosion in the quantity of data acquired during routine light microcopy. At the single-molecule level, analysis involves multiple steps and can rapidly become computationally expensive, in some cases intractable on office workstations. Complex bespoke software can present high activation barriers to entry for new users. Here, we redevelop our quantitative single-molecule analysis routines into an optimized and extensible Python program, with GUI and command-line implementations to facilitate use on local machines and remote clusters, by beginners and advanced users alike. We demonstrate that its performance is on par with previous MATLAB implementations but runs an order of magnitude faster. We tested it against challenge data and demonstrate its performance is comparable to state-of-the-art analysis platforms. We show the code can extract fluorescence intensity values for single reporter dye molecules and, using these, estimate molecular stoichiometries and cellular copy numbers of fluorescently-labeled biomolecules. It can evaluate 2D diffusion coefficients for the characteristically short single-particle tracking data. To facilitate benchmarking we include data simulation routines to compare different analysis programs. Finally, we show that it works with 2-color data and enables colocalization analysis based on overlap integration, to infer interactions between differently labelled biomolecules. By making this freely available we aim to make complex light microscopy single-molecule analysis more democratized.

**Graphical abstract**

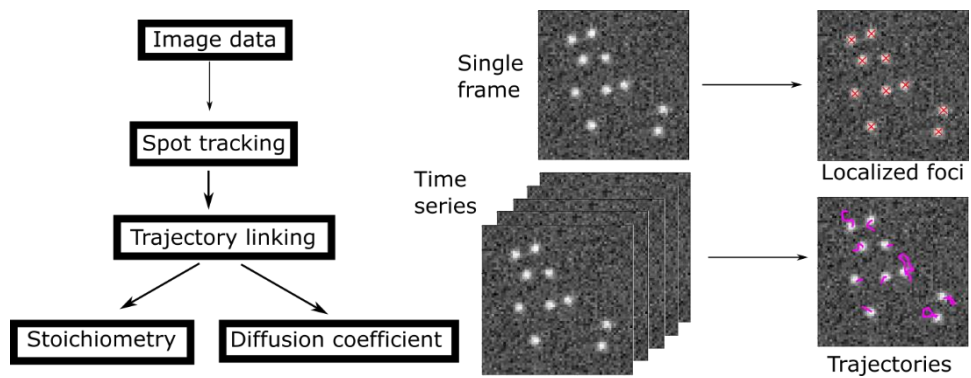

1. Introduction

Cell biology was transformed by the advent of super-resolution microscopy, a sub-theme of which is single-molecule localization microscopy (SMLM) [1]. SMLM techniques determine the spatial location of single fluorophores to below the optical diffraction limit by fitting a point spread function (PSF) to the experimentally acquired image data. These localizations can be used in a 'pointillist' method to reconstruct a single or time series super-resolved image, as in Photo-Activated Light Microscopy (PALM) [2] and Stochastic Optical Reconstruction Microscopy (STORM) [3], or single-molecules or clusters can be tracked as a function of time while quantifying their intensity and diffusion coefficients [4]–[7]. Particularly, analysis of intensity and step-wise photobleaching has become a powerful tool to measure the stoichiometry (i.e. the number of fluorescently labelled biomolecules present in any given tracked object) and copy number of molecular complexes in cells [8]–[14]. Multiple algorithms and software packages have been written and made available to researchers to analyze these super-resolution microscopy data either as standalone suites or as plugins for popular image analysis programs such as ImageJ [15]. However, limited software tools are available for stoichiometry determination and none are available, to our knowledge, exploiting the speed and extensibility of Python.

Existing super-resolution localization software has been extensively reviewed and compared [16], [17] but we discuss some of the more popular packages here. Among the most popular super-resolution reconstruction package is ThunderSTORM [18], a multi-purpose tool which is capable of reconstructing data from both STORM and PALM, techniques which both work to increase the temporal and spatial separation of emitting fluorophores so that the point spread function (usually approximated as a 2D Gaussian intensity profiles in the focal plane) can be fit to one fluorescence emitter only. ThunderSTORM is a powerful and flexible toolbox which gives high sub-pixel reconstruction accuracy, although for this to be the case the experiment must be optimized for and performed on fixed cells, and as a result dynamic information such as that embodied within effective diffusion coefficients are in general inaccessible. Similar approaches are also shared by other popular algorithms such as RainSTORM [19], QuickPALM [20] and DAOSTORM [21] which again produce high spatial resolution with the caveat that there is no temporal information. However, in the case of DAOSTORM, multiple point spread function fits allow the reconstructible density of fluorophores to rise by approximately sevenfold, while QuickPALM also includes utilities for 3D reconstruction and drift correction, processes that would generally be included in a larger multi-package workflow. Some routines have also been developed based not on classical algorithms but on machine learning in the case of 3B (standing for "Bayesian analysis of bleaching and blinking") [22], which hold the promise of more efficient analysis of large time-series data but which require careful interpretation of the results as well as considered choice of models and priors in the case of Bayesian statistics.

Away from STORM/PALM-type static reconstruction, many codes have been developed to find individual foci in noisy live-cell microscopy data. In general, classical algorithms in the same class as PySTACHIO and ADEMSCode operate through identification of local intensity maxima, though some include pre-filtering steps such as Gaussian filtering [23]–[27], Laplacian of Gaussian [25], [26], [28], wavelet products [29], [30], or deconvolution [31]. In general, a functional form is then fit to detected peaks (commonly Gaussian but occasionally Lorentzian[32] ), though in some cases localization itself is done using adaptive thresholding methods [27]. PySTACHIO and ADEMSCode both use Gaussian filtering, peak detection, intensity threshold masking, and finally iterative Gaussian fitting, meaning spot detection in PySTACHIO is comparable to state-of-the-art methods.

Having found spots in individual image stack frames, the challenge is then to compile these into individual focus trajectories. Here, PySTACHIO and ADEMSCode use the most conservative approach, which is to link spots between frames based on distance thresholding, as some other algorithms do [30], though some also include thresholding on the shape of the fitted Gaussian function to determine whether two foci are the same particle. However, more exotic algorithms are also in use today, such as multiple hypothesis tracking [33], probabilistic data association [34], and nearest-neighbor assignment [24]. Many of these also make use of so-called 'dropped frame' tolerance [17] – that is to say, if a spot exists in a position ($x,y$) in frame $n$, is not detected in frame $n+1$, but is localized near to ($x,y$) in frame $n+2$ the trajectory is accepted and the 'dropped' localization is filled in *a posteriori.* While this has been shown to work well in some systems, we use the conservative strict-linking method in PySTACHIO to avoid the risk of mis-linking in the highly crowded and diffusive live cell environment.

After tracking, many packages are available for post-processing either trajectories or spot intensities. Spot diffusion can be analyzed to extract physically relevant properties such as the diffusion coefficient, or to elucidate modes of motion – i.e. tethered, semi-tethered or free diffusion, for example by trajectory postprocessing with Single-Molecule Analysis by Unsupervised Gibbs sampling (SMAUG) [35] which uses a machine learning approach to undercover the diffusion states underlying the determined fluorophore trajectories. Similarly, Bayesian approaches may be used to identify single fluorophore bleaching steps to estimate stoichiometries [36]. However, these are generally used after the tracking and trajectory determination has taken place and are more accurately classified as post-processing packages.

In Python, some single-molecule tracking codes have been developed, trackpy is based on the commonly used Crocker and Greir algorithm [24] and recently TRAIT2D [37] has also been developed. However, these packages are not capable of molecular stoichiometry analysis. In this paper, we present PySTACHIO, a standalone single-molecule image analysis framework written in Python 3.8 and based on our original MATLAB (MathWorks) framework [38], that had been developed and improved from a range of earlier core algorithms implemented both in MATLAB [39] and LabVIEW [40], [41] (NI), but used a MATLAB version and libraries that gave improvements in computational speed through parallelization of key For Loop structures [8]. Given single-molecule photobleach image series, PySTACHIO tracks molecule positions detected in the focal plane of the fluorescence microscope as a function of time and calculates their stoichiometry and diffusion coefficients. It fits a kernel density function to the measured background-corrected intensities and produces an estimate of the fluorescence intensity denoted as Isingle, that corresponds to the characteristic brightness of a single fluorophore molecule integrated over all pixels in the central circular region of the PSF minus any contributions due to local background such as camera noise, sample autofluorescence and of fluorophores that are not in the focal plane but still contribute fluorescence detected by the camera detector. This Isingle estimate can be used alongside interpolation and model fits of the fluorophore photobleaching probability to give the initial fluorescence intensity to estimate the stoichiometries of detected fluorescence foci and estimate the total copy numbers of fluorescence emitter inside individual whole cells. It includes an easy to use GUI which is configured to be installable as a web hosted app (at the time of writing we have a demonstration instance available for public use) as well as a command-line tool which may be used to run PySTACHIO on batches of data on remote clusters. PySTACHIO is written to be both modular and extensible and we hope that this skeleton application will be further developed by us and others in the future.

## 2. Methods

The underlying principles of PySTACHIO are the same as those in our previous code [38]. In brief, the algorithm works by generating candidate fluorescent foci from the raw image using an optional Gaussian blur followed by a top-hat transformation to detect the background. The image is then binarized, with the threshold automatically determined from the peak of the pixel intensity histogram. A series of morphological opening and closing is used to determine candidate pixels associated with individual fluorescent foci. The center coordinates are then optimized through iterative Gaussian masking which when converged, reports the central position to sub-pixel accuracy with a precision related to the number of photons received from the fluorophore and the pixel size (a general rule of thumb for 5 ms exposure and a standard green fluorescent protein this lateral spatial precision is ~40 nm). Candidate foci are then assessed for signal-to-noise ratio (SNR) by comparing the integrated intensity within a 5 pixel radius of the candidate center coordinate with the standard deviation of the pixel intensities inside a larger 17x17 pixel square centered on the fluorescent focus center, excluding those within the center circle. Those that fall below the threshold (typically 0.4, whose value is informed by *in vitro* calibration data using surface immobilized fluorophores [10] combined with edge-preserving filters applied to the time-resolved data that allow single-molecule bleach steps to be detected directly [42]) are then removed from the candidate foci list, while the remaining accepted foci are then corrected for local background by subtraction of the mean of the intensities of the local background pixels within the 17x17 pixel square but excluding the 5 pixel radius circle.

Foci detected in successive frames are then linked into particle trajectories if the distance between them falls between a user-settable parameter, by default 5 pixels based around the typical width of the PSF, specifically approximately the full width at half maximum of a single GFP molecule PSF in our single-molecule microscope [43]. The linked foci are built up into a trajectory which is written to a file alongside key information at that frame – namely intensity, foci widths, and SNR values. These are trivially read in for post-processing or visualization either with PySTACHIO or with a range of bespoke software. If two trajectories collide, both are terminated at that frame at the coincident locus since this results in the lowest likelihood for incorrect linking of nearby fluorescent foci, but trivial user-modification of this criterion can enable linking-decision criteria based on physical parameters such as foci intensity to generate much longer trajectories if required [44].

Single-molecule foci intensities, Isingle, are estimated by taking the background-corrected intensities as calculated above for all foci, or optionally for all foci in the final half of the data acquisition in which most of the sample has been photobleached. The intensities are then binned into a histogram, and a kernel density function estimate (KDE) [12] fitted using the gaussian_kde routine from scipy with a kernel width set to 0.7 (set on the basis of typical estimates to size of Isingle compared to the background noise [45]). The peak of this fit is then found, and this is taken to be the Isingle value. Though we do not explicitly calculate or propagate errors on Isingle values (or other estimated values) an error bar may be estimated by taking the full width at half maximum value of the peak in the KDE plot which corresponds to Isingle. Note however that this approach relies on having good single-molecule data as an input to the routine – the data should for example be fairly low density, either monomeric fluorophores or photobleaching over the course of the acquisition. Once Isingle is found, it can be set as a parameter for future analysis runs rather than calculating it each time. Using the Isingle value, the molecular stoichiometry is found for each fluorescent focus by dividing its total integrated intensity by the Isingle value to give the value for the number of fluorophores present in that focus. For trajectories which begin in the first four frames of the acquisition, we fit a straight line to the first three intensity values of the trajectory and extrapolate back to the initial intensity, which is used to generate a stoichiometry value corrected for photobleaching. A linear fit is used as a compromise approximation to the expected exponential photobleach probability function, since it approximates the initial points of an exponential decay for higher stoichiometry foci to acceptable accuracy, but also fits the flat linear section of a step-wise photobleach of a lower stoichiometry fluorescent focus during which potentially no photobleaching may have occurred [46]. Other methods for stoichiometry determination involve counting the number of steps directly [47]. This works well for low copy number proteins in high SNR environments where single steps are easily resolved but is less general, although has been automated using methods such as Hidden Markov modelling [48].

Diffusion coefficients are generated from the detected trajectories by plotting the mean squared displacement as a function of time for each diffusing particle. The initial section of the mean squared displacement (MSD) vs. time interval relation for each tracked focus (by default, the first four time intervals values) is then fit with a straight line, and its gradient and intercept extracted. By default, the fitting algorithm constrains the intercept to be the known localization precision (this is a limitation of the current implementation – other work as demonstrated that in the presence of camera blur and other errors this assumption may be faulty [49]). The diffusion coefficient is then given as the gradient divided by four for 2D diffusion in the lateral focal plane of the microscope. Typically, trajectories of five frames or fewer are disregarded from the diffusion analysis, but this parameter may be modified by the user to account for longer or shorter duration trajectories depending on their specific imaging conditions.

Simulated diffusing and photobleaching fluorescent foci are created with an initially pseudo-random position. If the diffusion coefficient is non-zero, the fluorophore is assigned a pseudo-random displacement drawn from a distribution designed to give the input diffusion coefficient as time t→∞. The foci photobleach after a pseudo-random time, the scale of which is set by a user-set bleach time parameter. If the maximum stoichiometry is above 1 molecule, each initial fluorescent focus is given a pseudo-random number of fluorophores and hence has intensity n*Isingle. After each frame, each fluorophore has a probability of photobleaching and those that do have their brightness removed from the simulation while the others remain. This static probability of photobleaching on each frame mimics the step-wise photobleaching behavior of clusters of fluorophores and can be used for Isingle analysis (see Figure 2). Note that here that unlike state-of-the-art fluorescence simulation packages (e.g. FluoSim [50]) we do not seek to model exact fluorophore photophysics so parameters such as fluorescence lifetime, photoblinking, and emission distributions are neglected. Instead, in PySTACHIO the desired number of fluorophores are seeded in an "on" (or emitting) state, and stochastically photobleach with a user-settable probability per frame which leads to an overall exponential decay of emitters. After photobleaching, fluorophores do not return to the on state. Fluorophores photobleach with a uniform probability of photobleaching at any point within a frame exposure. To simulate this, we generate a uniform random number between 0 and 1 and give the following frame that fraction of Isingle in addition to the n*Isingle that it receives due to the emitters in the on state. During diffusion simulations, fluorophore movement occurs as a step at the end of each frame and the fluorophores are assumed to be static throughout the frame integration time – an assumption which significantly improves computational efficiency, but which could be improved in later version of the codebase. Similarly, we do not model fluorophores diffusing in and out of the plane of focus which would require not only 3D diffusion but also a 3D points spread function, increasing computational complexity considerably.

A graphical user interface (GUI) which runs locally in a browser window was written using plotly Dash and is capable of selecting files, running analysis, changing parameters, and showing results and simulated data on separate tabs. On the command line, we make use of Python 3's multiprocessing module to parallelize the tracking portion of the code using multiple CPU cores in a way analogous to OpenMP. PySTACHIO is not GPU-accelerated at this time.

The overall workflow of PySTACHIO is given in flowchart form in Figure 1a.

## 3. Results

*3.1 PySTACHIO performs well at identifying foci in simulated data*

Figure 1b shows simulated image data with crosses overlaid at the detected positions of simulated fluorophores, where the simulation parameters were taken to be consistent with experimentally observed values (Isingle=10,000 bg_mean=500 bg_std=120 num_spots=10 frame_size=(128,128) diffusion_coeff=1.0 pixel_size=0.120 [these are the default simulation parameters for both the installable PySTACHIO and the web-hosted instance]). By measuring detected positions and comparing to the known simulated ground truth, we can plot the root mean squared error (Figure 1c). We note that that these errors are sub-pixel in scale with the modal error being around 0.2 pixels, a distance in our simulation of approximately 20 nm, comparable to previous experimental findings [51]. In Figure 1b, we see that in this case out of ten spots with optimal parameter choices (snr_filter_cutoff=0.4 bw_threshold_tolerance=0.8 num_frames=2 subarray_halfwidth=8 inner_mask_radius=3 max_displacement=7 filter_image=Gaussian min_traj_len=2) all ten are detected, which is consistent with (though slightly superior to) previous detection accuracies with this method [38] – however, this is highly dependent on well-optimized parameter choices.

We have also applied PySTACHIO to previously generated challenge data [17] using the SNR=4 diffusing data set which was noted to be the threshold for most packages to reliable super-resolve spots. Run on single frames with optimal parameter choices (snr_filter_cutoff=0.6 num_frames=100 pixel_size=0.067 bw_threshold_tolerance=0.5 subarray_halfwidth=8 struct_disk_radius=10 inner_mask_radius=3 max_displacement=7 filter_image=Gaussian min_traj_len=3), we find that 83-100% of spots are identified, with an average detection rate 92%. Here, we used a radius cutoff of 2 pixels to discriminate between false and true positives. False positives range between 0 and 4 per frame with an average 1.3 false positive spots per frame (note that each simulated frame here has *ca.* 50 spots so this represents a low percentage error). Per frame, we find between 0 and 12 false negatives with an average of 5.5 false negatives per frame. This is consistent with PySTACHIO and ADEMSCode performance on other trial data – we find that in general false negatives outnumber false positives as spots are discarded which are too close together and cannot be found if they are too close to the frame edge, as the bounding box would then extend beyond the frame itself. With these detection and error rates, we report a frame-by-frame Jaccard similarity index 0.8-1.0, mean 0.91. Compared to the ground truth data, we find a root mean square localization error of 0.47 pixels, which at this simulated pixel size corresponds to approximately 30 nm.

However, PySTACHIO's more common operation mode is trajectory linking, and with this enabled we also discard any spots which are not part of a trajectory with a length greater than a user-specified cutoff (usually three frames). This leads to higher error rates but fewer false positives. Running PySTACHIO with trajectory linking reflects this. Here, we find an average true positive rate of 81.9 (range 66.1% to 92.5%), average false negatives per frame increase to an average of 14.1 false negatives per frame (range 6-22), and false positives reduce to an average of 0.5 false positives per frame (range 0-4), leading to an average Jaccard similarity index of 0.81 (range 0.65-0.93). We note here that we do not correct for putative 'dropped frames' as do other software platforms [17] – we insist on strict linking where each spot must be detected and localized within the cutoff radius for each frame step. In the highly diffusive subcellular environment this strict linking increases confidence in individual tracks though does so at the cost of removing some trajectories from later analysis.

We also used the challenge data to accurately measure the performance of our code compared to that of our previous version ADEMSCode. We found that with the same parameter set, PySTACHIO tracked all 100 frames in *ca.* 60 s while it took ADEMSCode around 560 s for the same tracking operation – a speedup in the new version of approximately 10x.

### 3.2 Simulating step-wise photobleaching

By giving each simulated fluorescent focus a notional number of fluorophores, we can simulate clusters of proteins. In the simulation parameters, we specify a probability of each fluorophore photobleaching between simulated frames. To simulate the next frame therefore we iterate through each fluorophore and generate a uniform pseudo-random number to determine if the fluorophore has photobleached (trivial modifications also allow users to define different probability distributions depending on the photophysics of the dye under study and the imaging environment). Repeating this for many frames gives an image where initially bright foci decay in a stochastic step-wise manner with an underlying exponential probability, as seen in Figure 2b. We have also implemented the Chung-Kennedy step-preserving filter [12] here which is shown as an inset to Figure 2b.

### 3.3 Single fluorophore brightness determination, and measuring stoichiometry

Tracking the intensity of all the foci across all frames we can form a histogram and approximate this with a Gaussian kernel density function with a specified bandwidth. By taking the peak of this KDE we approximate the underlying Isingle value, i.e., the integrated intensity of a single molecule (Figure 2a). Dividing the initial brightness of the focus, we can find the number of fluorophores that compose it, the so-called stoichiometry. We estimate the t=0 intensity of the focus by fitting the intensities of the focus in the second, third, and fourth frames with a straight line and extrapolating this back to the first frame to approximately correct for photobleaching. This extrapolated brightness is then divided by the Isingle value to give the stoichiometry. Testing this on simulated data gives excellent agreement with the input ground truth values (Figure 2c). It is easy to modify the form of the interpolation function as required, for example to use an exponential interpolation, however, a straight line we found to be a pragmatic compromise to both approximate a short section of an exponential photobleaching response function but also provide reasonable interpolation in instances where no photobleaching of track foci had actually occurred for which exponential interpolation would be unphysical.

### 3.4 Generating trajectories for simulated diffusing fluorophores

By comparing localized foci between frames and applying a distance threshold, we work out which pairs of foci are likely to be the same molecule. These have their positions linked between frames to form a trajectory. Comparing the input ground truth to the measured trajectory (Figure 3a) shows an excellent level of correspondence, with the same distribution of absolute errors as in Figure 1c.

### 3.5 Determining diffusion coefficients in simulated data

To determine the diffusion coefficient for each tracked fluorescent focus, we begin by plotting the MSD against time interval, τ (Figure 3b). According to Brownian motion, these plots should be a straight line whose gradient is four times the diffusion coefficient. We therefore fit a straight line and extract the gradient to estimate the diffusion coefficient. In order to avoid biases due to unusually long trajectories, by default we take only the first four MSD plot points, and we weight the linear fit to these towards the lower τ values containing more points. In our previous MATLAB implementation this was also constrained such that the intercept of the fit passed through the known localization precision. The default setting in PySTACHIO performs an unconstrained fit to cover instances where users have not measured the localization precision; however, we found that the average diffusion coefficient estimate is still within errors of the ground truth. As we see in Figure 3c the straight-line fits give a distribution of values centered around the simulated ground truth. Running and tracking ten simulations at each simulated diffusion coefficient, we build up statistics as in Figure 3d. Although the spreads are relatively high, the ground truth line hits each interquartile range which for single-molecule data is an acceptable level of accuracy. We note however that in general our estimations skew marginally lower than the ground truth values. We hypothesize this to be due to the step-length distributions in each simulation. As diffusion coefficient increases, the chance of a fluorophore moving a step length greater than our distance cutoff for a fluorophore to be linked between successive frames goes up. Because of this, trajectories may be split into two parts, each of which necessarily contains the lower-apparent-diffusion parts of the trajectory. Although this is a weakness, it is common to all distance-cutoff methods and underlines the need for thoughtful selection of parameters based on fluorophore density and the physical properties of the system under investigation. We also note that this small bias is in all case significantly less than the standard deviation.

### 3.6 PySTACHIO computational efficiency

Figure 4 shows the computational scaling of PySTACHIO with common variables. In Figure 4a, the scaling of PySTACHIO shows the expected quadratic scaling with frame size, though with an artefact for low frame sizes. These simulations were performed with a fixed number of simulated foci and as such, as the frame size increases the effective focus density is reduced. This is correlated with a decrease in overall runtime despite the larger frame. We hypothesize that in some circumstances Gaussian masking can take significantly longer to converge in the case that there are two or more fluorophores in close proximity that lead to heightened or irregular local backgrounds, leading to overall profiling of the Gaussian masking to get a higher standard deviation of runtime as shown in Supplementary Figure 1. Between the 64x64 and 128x128 pixel simulations therefore the higher overhead of the larger frame is outweighed by the cost savings of fluorophores which are more spatially separated.

In Figure 4b we see the scaling due to number of foci (though with a large enough frame size that the fluorophores remain spatially separated), while in Figure 4c the scaling due to number of frames. In each case the scaling is linear, which is the expected behavior given the O(N) scaling considerations in each case.

### 3.7 GUI and terminal modes

As well as being run in the terminal, plotly.dash was used to create a browser-based dashboard. Here, users can select files for tracking and post-processing and change key parameters to observe their effect on results. Users can also choose to simulate data within the GUI application and is therefore most suited to smaller datasets, new users, or exploratory/preliminary analysis.

By contrast, the terminal application supports batch processing and runs in headless mode with results written to files including graph generation for usual usage modes, such as stoichiometry calculation, diffusion coefficient calculation, and so on. Usage on the command line is in the following format: PySTACHIO.py tasks file_root keyword_args where tasks is one or more from track simulate postprocess view where the arguments must be separated by commas but without spaces; file_root is the path and root name of the file to be tracked (if in simulation mode, this is used for output files) and should be specified without the .tif extension. This root is used also for all the output files and plots. keyword_args allow the user to specify individual parameters to override defaults, e.g. snr=0.5. The command line implementation can therefore be trivially used to script convergence tests across a range of parameters, producing graphs for each condition.

*3.8 Visible copy number analysis*

If the user supplies a binary cell mask in .tif format where pixels of value 0 represent background, value 1 pixels belong to cell 1, PysSTACHIO will find the integrated and background-corrected intensity for each cell in the first bright frame and report an approximate copy number for that segmented binary large object (BLOB), valuable for users who wish to know how many fluorescently labelled biomolecules are, for example, present in any given single biological cell. Under tests (see Figure 5a) we simulated 100 fluorophores pseudo-randomly distributed in a 3D rod-like bacterial cell typical of many light microscopy investigations, focused at the midplane of the cell. We performed this ten times with varying noise. The mean total copy number was 99 ± 0.2(S.E.M.), once corrected for the presence of any of out-focal-plane fluorescence [51].

*3.9 Linking foci in dual-color experiments*

For two-color experiments, often employed to enable whether different biomolecules in a cell interact with each other, the color channels are analyzed separately initially as for single color microscopy. The tracked foci data for each position are used to generate the distances between each set of fluorophores between frames in each channel. Foci pairs with a distance higher than a user-settable cut-off (default five pixels) are discarded. The rest have an overlap integral calculated using their fitted Gaussian widths, and if this integral is above a threshold the pairs are taken to be colocalized [39]. In experimental data, such putative colocalization can then be indicative of binding between tagged molecules, at least to within the experimental localization precision of typically a few tens of nanometers.

Tests on simulated data (Figure 5b) show that the algorithm works well in high SNR regimes, with all located foci correctly linked. However, the simulated data has various simplifications not present in real data. First, simulated two color data has perfect registration between channels, while for real data channels can be misaligned or contain chromatic and other aberrations necessitating linear or affine transformation between channels and tracked foci data. Depending on the microscope this may introduce a significant source of error. In simulated data, the foci are high SNR and have the same SNR across colors which is generally not true for real life data and again introduces error. Careful interpretation of output data is therefore necessary.

*3.10    Comparison to live cell data*

We compared PySTACHIO to previously describe single-molecule localization data obtained from a study of a fluorescently labeled transcription factor, Mig1, inside live budding yeast cells [1] and analyzed trajectories for foci stoichiometries. Our results (Figure 5 panels c and c) show good agreement with previously described results. A fitted Gaussian kernel density estimation shows a peak at 4.4 which as half width at half maximum 4.5, a range which is within error of published results for a cluster size of associated Mig1 molecules [4], [46].

**4   Discussion**

Our single-molecule analysis software has been translated into Python and is now between 10 and 20x faster than the MATLAB implementation. It also has a user-friendly interface alongside a simple-to-script command line interface for power users. Our results work well on simulated data and are comparable to previous analyses of experimental data.

PySTACHIO is capable not only of tracking particles and track analysis but also simulation and molecular stoichiometry calculation for even high (10s-100s) stoichiometries. It is written entirely in Python 3.8 and free packages for Python and is written in a modular and extensible way to facilitate customization for a wide array of image analysis projects. PySTACHIO is released under the MIT license allowing anyone to download and modify our code at any time. We hope therefore that our program will be accessible for new users and democratize image analysis as well as forming a basis for advanced users to interrogate their data in depth. Particularly, there is enormous potential to integrate PySTACHIO into recent Python microscope control software [52], [53].

**Code availability** The PySTACHIO source is available to download from GitHub at https://github.com/ejh516/pystachio-smt. A static version of the code used for this publication is available via Zenodo [54]. PySTACHIO will soon be available as an installable package on PyPI as pystachio-smt. A web-hosted instance is available at the time of writing for public use which contains the key utilities of the code as described to enable users to explore its functionality prior to downloading locally and adapting to their own specific needs. Details of how to access this web version are available in the GitHub.

**CRediT author statement**

**JS:** Conceptualization, Data curation, Formal analysis, Investigation, Methodology, Software, Validation, Visualization, Writing - original draft, Writing - review & editing. **EH:** Data curation, Formal analysis, Investigation, Methodology, Software, Validation, Visualization, Writing - original draft, Writing - review & editing. **AW:** Methodology, Supervision, Validation, Writing - original draft, Writing - review & editing. **ML:** Conceptualization, Funding acquisition, Project administration, Resources, Supervision, Writing - original draft, Writing - review & editing.

**Acknowledgements** This work was supported by funding from the Leverhulme Trust (RPG-2019-156) and the Biotechnology and Biological Sciences Research Council BBSRC (BB/R001235/1). Many thanks to Emma Barnes (University of York IT services) for supporting the secondment of EJH to this project.

**Figures and captions**

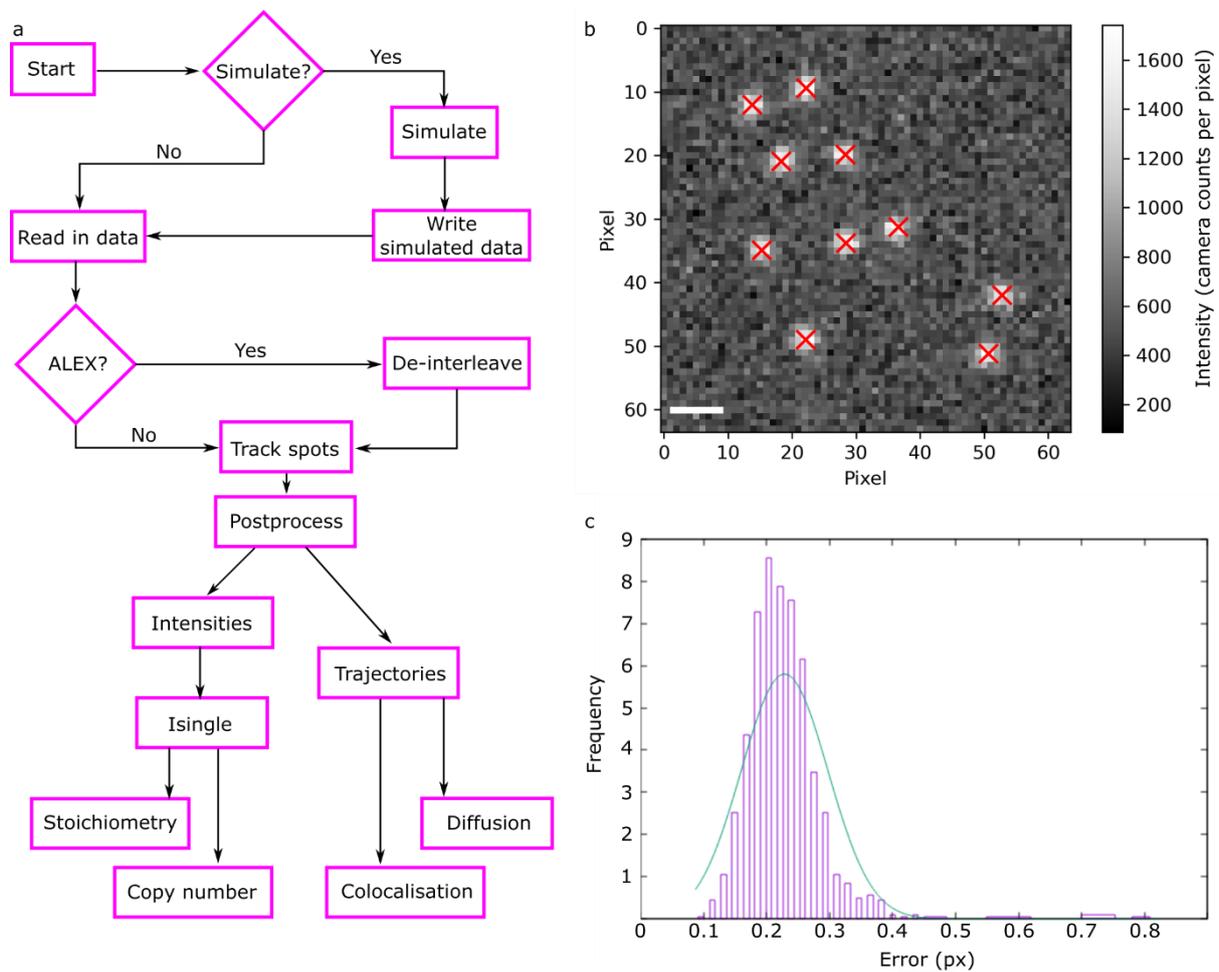

**Figure 1:** a) Flowchart of the PySTACHIO workflow; b) simulated data with identified foci indicated with red crosses. Here, the foci were simulated with Isingle 14,000, pixels were 120x120nm in size, and the background had mean and standard deviation 500 and 120 counts respectively. c) Error on simulated foci in pixel units. Bar: 1 μm.

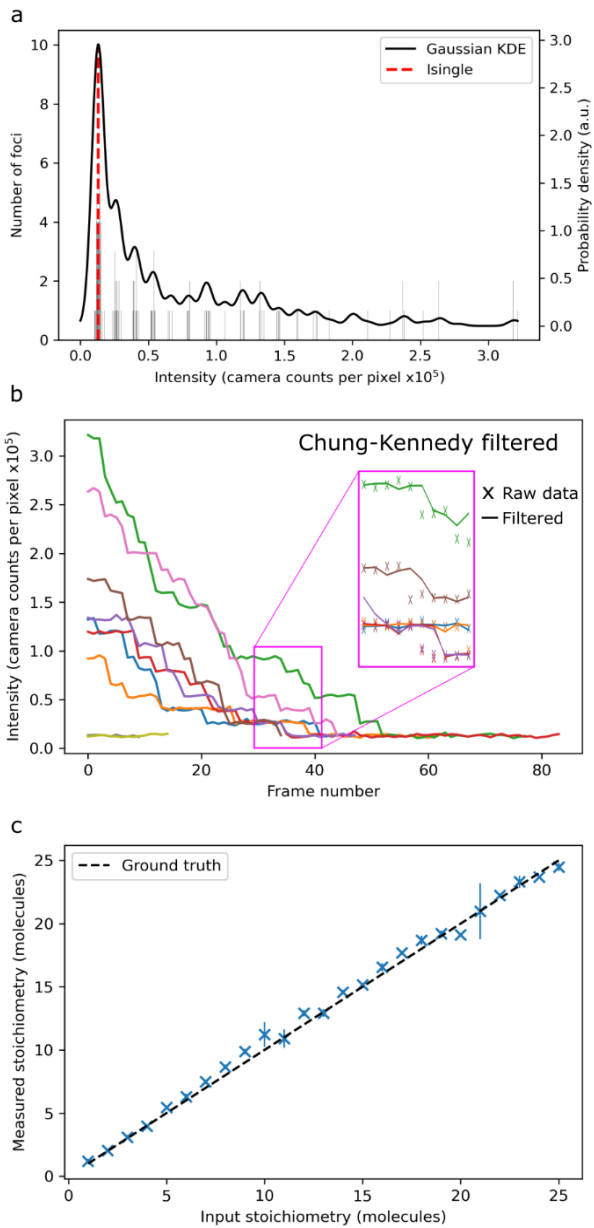

**Figure 2:** Simulated step-wise photobleaching of immobile multi-fluorophore foci. a) The KDE fit of measured intensities gives an accurate estimation of Isingle (input Isingle ~14,000 counts); b) intensity plots of the tracked foci show characteristic photobleaching steps. Inset: Chung-Kennedy [42] filtered intensity traces show clear steps; c) the rounded stoichiometry reproduces the input stoichiometry within error across the stoichiometry range 1-25 molecules.

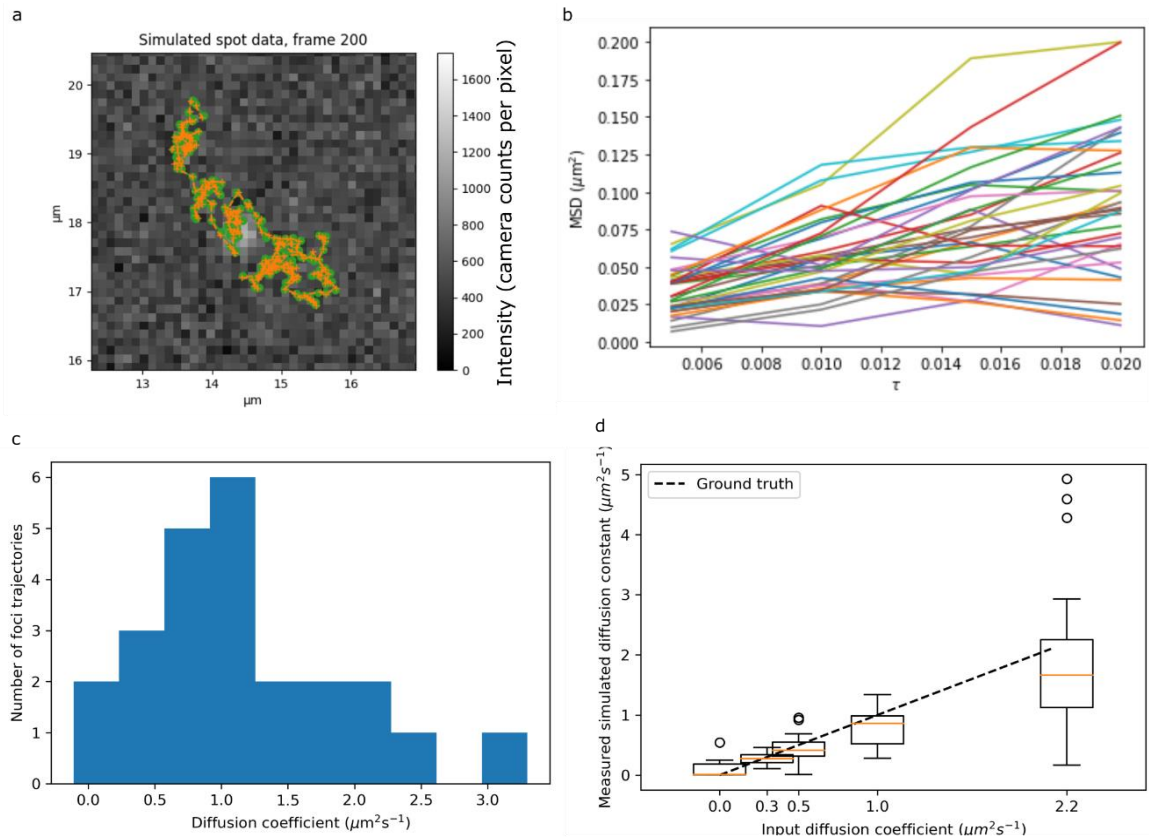

**Figure 3:** a) Simulated fluorophore trajectory with the tracked trajectory overlaid; b) mean squared displacement (MSD) plots for diffusing fluorophores; c) histogram of measured diffusion coefficients; d) box plot showing the distribution of measured diffusion coefficients for given input diffusion coefficients. Here the orange central line is the mean, with the box itself representing interquartile range (IQR). The whiskers represent the IQR ± one standard deviation, and circles show datapoints outside this range. In all cases, the ground truth line (dashed in black) passes through the interquartile range of the measured diffusion coefficients. The upper simulated limit for diffusion coefficient is set by theoretical considerations of the maximum detectable diffusion coefficient based on the criterion of a maximum of a five pixel separation between foci in subsequent image frames to be considered part of the same focus trajectory assuming rapid Slimfield millisecond single-molecule microscopy [55].

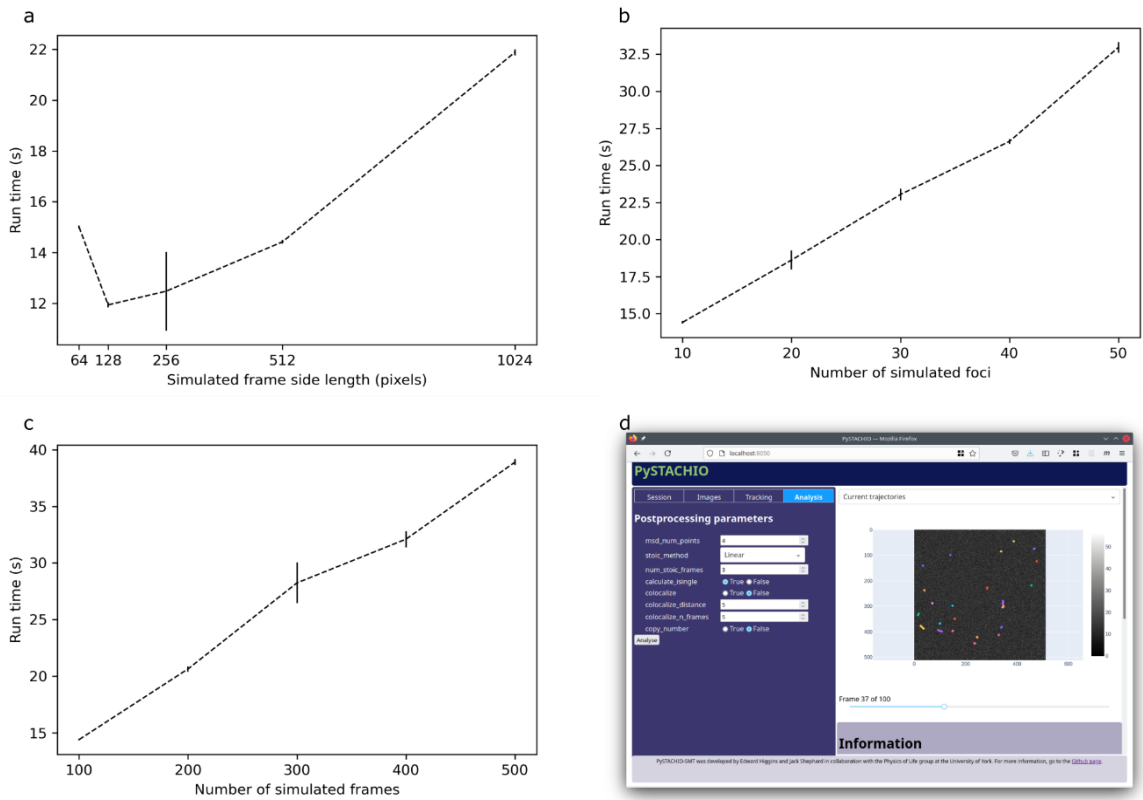

**Figure 4:** Scaling of runtime for PySTACHIO with a) frame size, b) kinetic series length, and c) number of foci to track; d) a screenshot from the GUI mode showing parameter selection and tracked trajectories. In panels a-c the error bars represent standard deviation. For each data point, the tracking software was run five times. In panels b) and c) frame size was 256x256 pixels. In panels a) and c) the number of simulated foci was 10. In panels a) and b) 100 frames were simulated.

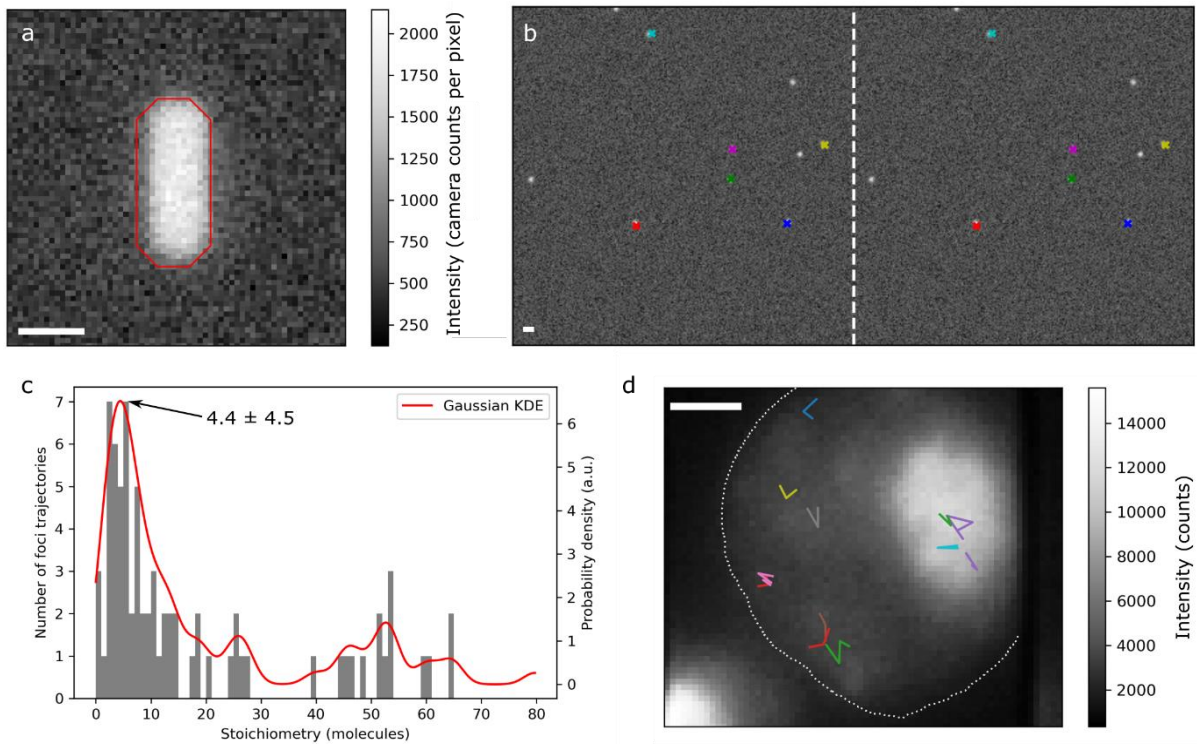

**Figure 5:** a) Simulated rod-like cell with red outline indicated specified mask used for copy number analysis; b) colocalized foci in a 2-colour experiment (simulated ALEX data here presented de-interleaved for clarity). Colocalized foci are indicated by the same color in both channels. The border between the left hand and right hand channel is indicated by a vertical dashed white line; c) stoichiometries taken from live-cell data in good agreement with previously published values, with peak stoichiometry 4.4±4.5 molecules; d) trajectories determined from the live-cell data overlaid on the mean of the five first bright fluorescence frames of the acquisition. The approximate cell outline is shown with a white dotted line. All scale bars: 1 µm.